\documentclass[12pt]{article}
\usepackage{amssymb}
\begin{document}



\def\a{\alpha}
\def\b{\beta}
\def\d{\delta}
\def\e{\epsilon}
\def\g{\gamma}
\def\h{\mathfrak{h}}
\def\k{\kappa}
\def\l{\lambda}
\def\o{\omega}
\def\p{\wp}
\def\r{\rho}
\def\t{\tau}
\def\s{\sigma}
\def\z{\zeta}
\def\x{\xi}
\def\V={{{\bf\rm{V}}}}
 \def\A{{\cal{A}}}
 \def\B{{\cal{B}}}
 \def\C{{\cal{C}}}
 \def\D{{\cal{D}}}
\def\G{\Gamma}
\def\K{{\cal{K}}}
\def\O{\Omega}
\def\R{\bar{R}}
\def\T{{\cal{T}}}
\def\L{\Lambda}
\def\f{E_{\tau,\eta}(sl_2)}
\def\E{E_{\tau,\eta}(sl_n)}
\def\Zb{\mathbb{Z}}
\def\Cb{\mathbb{C}}

\def\R{\overline{R}}

\def\beq{\begin{equation}}
\def\eeq{\end{equation}}
\def\bea{\begin{eqnarray}}
\def\eea{\end{eqnarray}}
\def\ba{\begin{array}}
\def\ea{\end{array}}
\def\no{\nonumber}
\def\le{\langle}
\def\re{\rangle}
\def\lt{\left}
\def\rt{\right}

\newtheorem{Theorem}{Theorem}
\newtheorem{Definition}{Definition}
\newtheorem{Proposition}{Proposition}
\newtheorem{Lemma}{Lemma}
\newtheorem{Corollary}{Corollary}
\newcommand{\proof}[1]{{\bf Proof. }
        #1\begin{flushright}$\Box$\end{flushright}}

\baselineskip=20pt

\newfont{\elevenmib}{cmmib10 scaled\magstep1}
\newcommand{\preprint}{
   \begin{flushleft}
   \end{flushleft}\vspace{-1.3cm}
   \begin{flushright}\normalsize
   \end{flushright}}
\newcommand{\Title}[1]{{\baselineskip=26pt
   \begin{center} \Large \bf #1 \\ \ \\ \end{center}}}
\newcommand{\Author}{\begin{center}
   \large \bf
Yuan-Yuan Li${}^{a}$,~Junpeng Cao${}^{a}$,~Wen-Li Yang${}^{b,c}$,~Kangjie Shi${}^b$ and~Yupeng
Wang${}^a\footnote{Corresponding author: yupeng@iphy.ac.cn}$
 \end{center}}
\newcommand{\Address}{\begin{center}

     ${}^a$Beijing National Laboratory for Condensed Matter
           Physics, Institute of Physics, Chinese Academy of Sciences, Beijing
           100190, China\\
     ${}^b$Institute of Modern Physics, Northwest University,
     Xian 710069, China\\
     ${}^c$Beijing Center for Mathematics and Information Interdisciplinary Sciences, Beijing, 100048,  China
   \end{center}}
\newcommand{\Accepted}[1]{\begin{center}
   {\large \sf #1}\\ \vspace{1mm}{\small \sf Accepted for Publication}
   \end{center}}

\preprint
\thispagestyle{empty}
\bigskip\bigskip\bigskip

\Title{Exact solution of the one-dimensional Hubbard model with
arbitrary boundary magnetic fields  } \Author

\Address
\vspace{1cm}

\begin{abstract}
The one-dimensional Hubbard model with arbitrary boundary magnetic fields is
solved exactly via the Bethe ansatz methods. With the coordinate
Bethe ansatz in the charge sector, the second eigenvalue problem
associated with the spin sector is constructed. It is shown that the
second eigenvalue problem can be transformed into that of the inhomogeneous
XXX spin chain with arbitrary boundary fields which can be solved
via the off-diagonal Bethe ansatz method.

\vspace{1truecm} \noindent {\it PACS}: 71.10.Fd; 75.10.-b; 02.30.Ik



\end{abstract}

\newpage
\section{Introduction}
The Hubbard model is one of the essential models in condensed matter
physics. An interesting issue is that the model is exactly solvable
in one dimension \cite{lieb-wu}, which provides an important
benchmark for understanding the Mott insulators. After Lieb and Wu's
pioneering work, a lot of attentions have been paid to the
integrability, symmetry \cite{Yan89,Yan90,Per90,Ess91} and physical
properties of this model. A remarkable result was obtained by
Shastry \cite{sha} and by Olmedilla et al \cite{Olm87,Olm88} who
constructed the corresponding $R$-matrix of the one-dimensional
Hubbard model and therefore demonstrated its complete integrability
in the framework of Yang-Baxter equation \cite{yang,bax}.
Subsequently, the model was resolved \cite{matin} via the algebraic
Bethe ansatz method based on the result of Shastry. Another
interesting issue about this model is the open-boundary problem,
which is tightly related to the impurity problem in a Luttinger
liquid \cite{wang}. The exact solution of the open Hubbard chain was
firstly obtained by Shulz \cite{shul}. Subsequently, the exact
solution of the model with boundary potentials was obtained
\cite{zhou,suzu}. The integrability of the one-dimensional Hubbard
model with diagonal open boundary was demonstrated in \cite{guan1}
by constructing the Lax representation and solved by algebraic Bethe
ansatz method \cite{guan2}. The generic integrable boundary
conditions were obtained in \cite{wada} by solving the reflection
equation \cite{skl,nepo,Kor93}. It was found \cite{wada} that in the
spin sector magnetic fields applied on the two end sites do not
break the integrability of this model. Although the integrability
has been known for long time, the  exact solutions (or
diagonalization of the Hamiltonian)  of the model with arbitrary
boundary magnetic fields are still lacking.

In this paper, we study the open Hubbard chain with arbitrary
boundary magnetic fields. The Hamiltonian of the model  is
\begin{eqnarray}
H&=&-t\sum_{\alpha,j=1}^{N-1}[c_{j,\alpha}^\dagger
c_{j+1,\alpha}+c_{j+1,\alpha}^\dagger
c_{j,\alpha}]+U\sum_{j=1}^Nn_{j,\uparrow}n_{j,\downarrow}
+h_1^{-}c^{\dagger}_{1,\uparrow}c_{1,\downarrow}
+h_1^+c^{\dagger}_{1,\downarrow}c_{1,\uparrow}\no\\
&&\quad +
h_1^z(n_{1,\uparrow}-n_{1,\downarrow})
+h_N^{-}c^{\dagger}_{N,\uparrow}c_{N,\downarrow}+h_N^
+c^{\dagger}_{N,\downarrow}c_{N,\uparrow}+
h_N^z(n_{N,\uparrow}-n_{N,\downarrow}),\label{Hamiltonian}
\end{eqnarray}
where $c_{j,\alpha}^\dagger$ and $c_{j,\alpha}$ are the creation and
annihilation operators of electrons on site $j$ with spin component
$\alpha=\uparrow,\downarrow$; $t$ and $U$ are the hopping constant
and the on-site repulsion constant as usual;
$n_{j,\alpha}$ are particle number operators,
respectively; ${\vec h}_1=(h_1^x,h_1^y,h^z_1)$ and ${\vec h}_N=(h_N^x,h_N^y,h^z_N)$ indicate the boundary
fields and $h^{\pm}_j=h^x_j\pm ih_j^y$ for $j=1,N$. We shall show in the following  that the model can be
exactly solved by combining the coordinate Bethe ansatz and the off-diagonal Bethe
ansatz proposed recently in \cite{cysw1,cysw2,cysw3} for arbitrary ${\vec h}_1$ and ${\vec h}_N$. We
remark that the unparallel boundary fields break the $U(1)$ symmetry in spin sector
and make the total spin no longer a conserved charge.

The paper is organized as follows. In Section 2, we use
the coordinate Bethe ansatz method to derive the eigenvalue equation
in the spin sector as that in the periodic case \cite{yang}. In Section 3, we
transform this eigenvalue problem into that of the inhomogeneous XXX spin
chain  with boundary fields, which allows us to apply the recently proposed
off-diagonal Bethe ansatz method \cite{cysw1,cysw2,cysw3} to solve it. The
exact spectrum of the Hamiltonian and the Bethe ansatz equations are
thus obtained. Section 4 is attributed to the reduction to the parallel or anti-parallel
boundary case. Concluding remarks are given in Section 5.

\section{Coordinate Bethe ansatz}
\label{CBA} \setcounter{equation}{0}
Though the $U(1)$ symmetry in the spin sector is broken by the
unparallel boundary fields, the $U(1)$ symmetry in the charge sector  
is still reserved. The conserved charge corresponding to this reserved symmetry is the total number operator of  electrons, 
namely, 
\begin{eqnarray}
  \hat{N}=\sum_{j=1}^N\left\{n_{j,\uparrow}+n_{j,\downarrow}\right\},\quad [H,\hat{N}]=0.
\end{eqnarray}
The symmetry allows us to construct the eigenstate of the Hamiltonian (\ref{Hamiltonian}) with a fixed number of electrons as follows:
\begin{eqnarray}
|\Psi\rangle=\sum_{j=1}^M\sum_{\alpha_j=\uparrow,\downarrow}\sum_{x_j=1}^N\Psi^{\{\alpha\}}(x_1,\ldots,x_M)
c_{x_1,\alpha_1}^\dagger\cdots c_{x_M,\alpha_M}^\dagger|0\rangle,
\end{eqnarray}
where $M$ is the number of electrons and
$\{\alpha\}=(\alpha_1,\ldots,\alpha_M)$. The eigenvalue equation of
the Hamiltonian then reads
\begin{eqnarray}
&&-t\sum_{j=1}^M[(1-\delta_{x_j,N})\Psi^{\{\alpha\}}(\ldots,x_j+1,\ldots)+(1-\delta_{x_j,1})\Psi^{\{\alpha\}}(\ldots,x_j-1,\ldots)]\no\\
&&\quad+U\sum_{i<j}^M\d_{x_i,x_j}\d_{\a_i,-\a_j}\Psi^{\{\alpha\}}(x_1,\ldots,x_M)\nonumber\\
&&\quad+\sum_{j=1}^M\delta_{x_j,1}[h_1^-\delta_{\alpha_j,\downarrow}\Psi^{(\ldots,-\alpha_j,\ldots)}(x_1,\ldots,x_M)
+h_1^+\delta_{\alpha_j,\uparrow}\Psi^{(\ldots,-\alpha_j,\ldots)}(x_1,\ldots,x_M)\no\\
&&\hspace{2.5cm}+h_1^z(\delta_{\alpha_j,\uparrow}-\delta_{\alpha_j,\downarrow})\Psi^{\{\alpha\}}(x_1,\ldots,x_M)]\no\\
&&\quad+\sum_{j=1}^M\delta_{x_j,N}[h_N^-\delta_{\alpha_j,\downarrow}\Psi^{(\ldots,-\alpha_j,\ldots)}(x_1,\ldots,x_M)
+h_N^+\delta_{\alpha_j,\uparrow}\Psi^{(\ldots,-\alpha_j,\ldots)}(x_1,\ldots,x_M)\no\\
&&\hspace{2.5cm}+h_N^z(\delta_{\alpha_j,\uparrow}-\delta_{\alpha_j,\downarrow})\Psi^{\{\alpha\}}(x_1,\ldots,x_M)]\nonumber\\
&&=E\,\Psi^{\{\alpha\}}(x_1,\ldots,x_M).
\end{eqnarray}
This eigenvalue equation can be rewritten as
\begin{eqnarray}
&&-t\sum_{j=1}^M[(1-\delta_{x_j,N})\Psi^{\{\alpha\}}(\ldots,x_j+1,\ldots)+(1-\delta_{x_j,1})\Psi^{\{\alpha\}}(\ldots,x_j-1,\ldots)]\no\\
&&\quad+U\sum_{i<j}^M\d_{x_i,x_j}\d_{\a_i,-\a_j}\Psi^{\{\alpha\}}(x_1,\ldots,x_M)\nonumber\\
&&\quad+\sum_{j=1}^M\sum_{\beta_j=\uparrow,\downarrow}}[\delta_{x_j,1}{\vec
h}_1\cdot{\vec \sigma_{\alpha_j,\beta_j}+\delta_{x_j,N}{\vec
h}_N\cdot{\vec
\sigma_{\alpha_j,\beta_j}}]\Psi^{\{\alpha\}_j}(x_1,\ldots,x_M)\nonumber\\
&&=E\,\Psi^{\{\alpha\}}(x_1,\ldots,x_M),\label{eigen1}
\end{eqnarray}
where $\{\alpha\}_j$ means $\alpha_j$ is replaced by $\beta_j$ in
the set $\{\alpha\}$ and $\vec \sigma=(\sigma^x,\sigma^y,\sigma^z)$ with $\sigma^{x}$,
$\sigma^{y}$ and $\sigma^{z}$ being the Pauli matrices. The  wave function takes the following Bethe
ansatz form \cite{yang}:
\begin{eqnarray}
\Psi^{\{\alpha\}}(x_1,\ldots,x_M)=\sum_{P,Q,r}A_P^{\{\alpha\},r}(Q)\exp[i\sum_{j=1}^Mr_{P_j}k_{P_j}x_{Q_j}]\theta(x_{Q_1}<x_{Q_2}<\ldots<x_{Q_M}),
\label{Wave-function}
\end{eqnarray}
where $P=(P_1,\ldots,P_M)$ and $Q=(Q_1,\ldots,Q_M)$ are the
permutations of $(1,\ldots,M)$; $r=(r_1,\ldots,r_M)$ with $r_j=\pm$
and $\theta(x_1<\ldots<x_M)$ is the generalized step function.  For
all $x_j\neq 1,N$ and $x_j\neq x_l$ case, (\ref{eigen1}) is
automatically satisfied and the corresponding eigenvalue is
\begin{eqnarray}
E=-2t\sum_{j=1}^M\cos k_j.\label{Eigenvalues}
\end{eqnarray}
For two electrons occupy the same site case, we should consider the continuity of the wave function $\Psi^{\{\alpha\}}(x_1,\ldots,x_M)$. Considering the sector $\uppercase\expandafter{\romannumeral1}$: $x_{Q_1}<x_{Q_2}<\ldots <x_{Q_j}<x_{Q_{j+1}}<\ldots <x_{Q_M}$ and the sector $\uppercase\expandafter{\romannumeral2}$: $x_{Q_1}<x_{Q_2}<\ldots <x_{Q_{j+1}}<x_{Q_j}<\ldots <x_{Q_M}$,
when $x_{Q_j}=x_{Q_{j+1}}=x$, the continuity of the wave function  $\Psi^{\{\alpha\}}(x_1,\ldots,x_M)$
demands
\begin{eqnarray}
\Psi_{\uppercase\expandafter{\romannumeral1}}^{\{\alpha\}}(\ldots,x,x,\ldots)=\Psi_{\uppercase\expandafter{\romannumeral2}}^{\{\alpha\}}(\ldots,x,x,\ldots).\label{continuity}
\end{eqnarray}
For convenience, we omit the superscript $\{\alpha\}$ and treat
$A_P^r(Q)$ as a column vector in the spin space. Then the continuity condition (\ref{continuity}) of the wave function implies
\begin{eqnarray}
A_P^r(Q)+A_{P'}^{r'}(Q)=A_P^r(Q')+A_{P'}^{r'}(Q'),\label{c1}
\end{eqnarray}
where $Q'=(\ldots,Q_{j+1},Q_j,\ldots)$,
$P'=(\ldots,P_{j+1},P_j,\ldots)$ and
$r'=(\ldots,r_{j+1},r_j,\ldots)$.
For $x_{Q_j}=x_{Q_{j+1}}\neq 1,N$, the Schr{\"o}dinger
equation (\ref{eigen1}) gives
\begin{eqnarray}
&&-t\lt[A_P^r(Q')e^{ir_{P_{j+1}}k_{P_{j+1}}}+A_{P'}^{r'}(Q')e^{ir_{P_j}k_{P_j}}+A_P^r(Q)e^{-ir_{P_j}k_{P_j}}+A_{P'}^{r'}(Q)e^{-ir_{P_{j+1}}k_{P_{j+1}}}\rt.\nonumber\\
&&\quad\quad +\lt.A_P^r(Q)e^{ir_{P_{j+1}}k_{P_{j+1}}}+A_{P'}^{r'}(Q)e^{ir_{P_j}k_{P_j}}+A_P^r(Q')e^{-ir_{P_j}k_{P_j}}+A_{P'}^{r'}(Q')e^{-ir_{P_{j+1}}k_{P_{j+1}}}\rt]\nonumber\\
&&\quad +U[A_P^r(Q)+A_{P'}^{r'}(Q)]=-2t[\cos k_{P_j}+\cos k_{P_{j+1}}][A_P^r(Q)+A_{P'}^{r'}(Q)].\label{2}
\end{eqnarray}
Substituting (\ref{c1}) into (\ref{2}), we have
\begin{eqnarray}
[\sin (r_{p_j}k_{p_j})-\sin (r_{p_{j+1}}k_{p_{j+1}})]A_P^r(Q)-i\frac U{2t}A_P^r(Q')\nonumber\\
=[\sin (r_{p_j}k_{p_j})-\sin (r_{p_{j+1}}k_{p_{j+1}})+i\frac U{2t}]A_{P'}^{r'}(Q').\label{c2}
\end{eqnarray}
Now, we define the coordinate permutation
operator ${\bar P}_{i,j}$,
\begin{eqnarray}
{\bar P}_{i,j}A_P^r(\ldots,x_{Q_i},\ldots,x_{Q_j},\ldots)=A_P^r(\ldots,x_{Q_j},\ldots,x_{Q_i},\ldots).
\end{eqnarray}
Due to the fact that the  wave function of fermion is completely antisymmetric under
exchanging both the coordinates and spins of two particles, if we denote
$P_{i,j}$ as the spin permutation operator, we have
\begin{eqnarray}
P_{i,j}{\bar P}_{i,j}=-1,\quad P_{i,j}^2={{\bar P}_{i,j}}^2=1.
\end{eqnarray}
Thus, we have the following relation:
\begin{eqnarray}
-P_{j,j+1}A_P^r(Q)=A_P^r(Q').
\end{eqnarray}
Substituting this relation into (\ref{c2}), we readily  have
\begin{eqnarray}
A_P^r(Q)=S_{P_j,P_{j+1}}(r_{P_j}k_{P_j},r_{P_{j+1}}k_{P_{j+1}})A_{P'}^{r'}(Q'),\label{s}
\end{eqnarray}
with the $S$-matrix given by
\begin{eqnarray}
S_{j,l}(k_j,k_l)=\frac{\sin k_j-\sin k_l-i\frac U{2t}P_{j,l}}{\sin
k_j-\sin k_l-i\frac U{2t}}.
\end{eqnarray}
Now let us turn to the case of $x_{Q_1}=1$, $x_{Q_i}\neq x_{Q_j}(i\neq j)$ and $x_{Q_M}\neq N$. In this case, the eigenvalue equation (\ref{eigen1}) becomes
\begin{eqnarray}
-t\Psi^{\{\alpha\}}(2,\ldots)+\sum_{\beta_1}{\vec
h}_1\cdot{\vec
\sigma}_{\alpha_1,\beta_1}\Psi^{(\beta_1,\ldots)}(1,\ldots)=-2t\cos k_{P_1}\Psi^{\{\alpha\}}(1,\ldots).
\end{eqnarray}
This induces
\begin{eqnarray}
\sum_{\beta_1}{\vec h}_1\cdot{\vec
\sigma}_{\alpha_1,\beta_1}\Psi^{(\beta_1,\ldots)}(1,\ldots)=-t\Psi^{\{\alpha\}}(0,\ldots),\label{B}
\end{eqnarray}
which gives
\begin{eqnarray}
{A_P^{(+,\ldots)}(Q)}&&=-[t+{\vec h}_1\cdot{\vec
\sigma_1}e^{ik_{P_1}}]^{-1}[t+{\vec h}_1\cdot{\vec
\sigma_1}e^{-ik_{P_1}}]A^{(-,\ldots)}_P(Q)\nonumber\\
&&\stackrel{def}{=}{\bar K}_1^+(k_{P_1})A_P^{(-,\ldots)}(Q).\label{k+}
\end{eqnarray}
With the help of the identity
\begin{eqnarray}
({\vec h}_1\cdot{\vec \sigma})^2={\vec h}_1^2,\nonumber
\end{eqnarray}
we have
\begin{eqnarray}
{\bar K}_j^{+}(k)=-\frac{t^2-{\vec h}_1^2-2it\sin k\,{\vec
h}_1\cdot{\vec \sigma}_j}{t^2-{\vec h}_1^2e^{2ik}}.
\end{eqnarray}
Similarly, for the case of $x_{Q_M}=N$, $x_{Q_i}\neq x_{Q_j}(i\neq j)$ and $x_{Q_1}\neq 1$, we have
\begin{eqnarray}
-t\Psi^{\{\alpha\}}(\ldots,N-1)+\sum_{\beta_M}{\vec
h}_N\cdot{\vec
\sigma}_{\alpha_M,\beta_M}\Psi^{(\ldots,\beta_M)}(\ldots,N)=-2t\cos k_{P_M}\Psi^{\{\alpha\}}(\ldots,N),
\end{eqnarray}
namely,
\begin{eqnarray}
\sum_{\beta_M}{\vec h}_N\cdot{\vec
\sigma}_{\alpha_M,\beta_M}\Psi^{(\ldots,\beta_M)}(\ldots,N)=-t\Psi^{\{\alpha\}}(\ldots,N+1),
\end{eqnarray}
which induces
\begin{eqnarray}
{e^{-2ik_{P_M}N}A^{(\ldots,-)}_P(Q)}&&=-[te^{-ik_{P_M}}+{\vec h}_N\cdot{\vec\sigma}_M]^{-1}[te^{ik_{P_M}}+{\vec h}_N\cdot{\vec\sigma}_M]A^{(\ldots,+)}_P(Q)\nonumber\\
&&\stackrel{def}={\bar K}_M^-(k_{P_M})A^{(\ldots,+)}_P(Q),\label{k-}
\end{eqnarray}
with
\begin{eqnarray}
{\bar K}_j^-(k)=-\frac{t^2-{\vec h}_N^2-2it\sin k\,{\vec h}_N\cdot{\vec
\sigma}_j}{t^2e^{-2ik}-{\vec h}_N^2}.
\end{eqnarray}
When $x_{Q_1}=x_{Q_2}=1$ or $x_{Q_{M-1}}=x_{Q_M}=N$, the situation
becomes a little bit subtle. We have to check the self-consistence
of the ansatz. For the case of $x_{Q_1}=x_{Q_2}=1$, $x_{Q_i}\neq x_{Q_j}(i\neq j$ and $i,j\neq 1)$ and $x_{Q_M}\neq N$, the eigenvalue equation (\ref{eigen1}) becomes
\begin{eqnarray}
&&-t[\Psi^{\{\alpha\}}(2,1,\ldots)+\Psi^{\{\alpha\}}(1,2,\ldots)]+U\Psi^{\{\alpha\}}(1,1,\ldots)\no\\
&&\quad+\sum_{\beta_1,\beta_2}[{\vec
h}_1\cdot{\vec \sigma}_{\alpha_1,\beta_1}+{\vec h}_1\cdot{\vec
\sigma}_{\alpha_2,\beta_2}]\Psi^{(\beta_1,\beta_2,\ldots)}(1,1,\ldots)\no\\
&&=-2t[\cos k_{P_1}+\cos
k_{P_2}]\Psi^{\{\alpha\}}(1,1,\ldots).\label{sub-2}
\end{eqnarray}
And in this case, the $S$-matrix makes the following equation hold:
\begin{eqnarray}
&&-t[\Psi^{\{\alpha\}}(2,1,\ldots)+\Psi^{\{\alpha\}}(0,1,\ldots)
+\Psi^{\{\alpha\}}(1,2,\ldots)+\Psi^{\{\alpha\}}(1,0,\ldots)]+U\Psi^{\{\alpha\}}(1,1,\ldots)\no\\
&&\quad=-2t[\cos k_{P_1}+\cos k_{P_2}]\Psi^{\{\alpha\}}(1,1,\ldots).\label{sub-3}
\end{eqnarray}
Combining Eq.(\ref{sub-2}) and Eq.(\ref{sub-3}), we have the following relation need to be confirmed:
\begin{eqnarray}
-t[\Psi^{\{\alpha\}}(0,1,\ldots)+\Psi^{\{\alpha\}}(1,0,\ldots)]=\sum_{\beta_1,\beta_2}[{\vec
h}_1\cdot{\vec \sigma}_{\alpha_1,\beta_1}+{\vec h}_1\cdot{\vec
\sigma}_{\alpha_2,\beta_2}]\Psi^{(\beta_1,\beta_2,\ldots)}(1,1,\ldots).\label{B3}
\end{eqnarray}
For the case of $x_{Q_1}=1$, $x_{Q_i}\neq x_{Q_j}(i\neq j)$ and $x_{Q_M}\neq N$, we have the relation (\ref{B}). For $x_{Q_2}=1$, $x_{Q_i}\neq x_{Q_j}(i\neq j)$ and $x_{Q_M}\neq N$, similarly, we have
\begin{eqnarray}
\sum_{\beta_2}\vec h_1\cdot\vec\sigma_{\alpha_2,\beta_2}\Psi^{(\alpha_1,\beta_2,\ldots)}(1,1,\ldots)=-t\Psi^{\{\alpha\}}(1,0,\ldots).\label{B2}
\end{eqnarray}
Obviously (\ref{B}) and (\ref{B2}) make (\ref{B3}) hold. With the same procedure
we can demonstrate that the ansatz is also satisfied when two electrons both occupy
the site $N$.

Now let us consider the following process. The $j$-th particle moves
from the $l$-th site to the left end by scattering with all the other
particles to their left, and then is reflected by the left boundary. After scattering with all the other particles, it
 is reflected by the right boundary and then moves back to its original position. This process can be described by the following relations:
\begin{eqnarray}
A^{(\ldots,+,\ldots)}&=&S_{j-1,j}(k_{j-1},k_j)S_{j-2,j}(k_{j-2},k_j)\cdots
S_{1,j}(k_1,k_j)A^{(+,\ldots)},\nonumber\\
A^{(+,\ldots)}&=&\bar{K}^+_j(k_j)\,A^{(-,\ldots)},\no\\
A^{(-,\ldots)}&=&S_{j,1}(-k_{j},k_1)\cdots S_{j,j-1}(-k_{j},k_{j-1})S_{j,j+1}(-k_{j},k_{j+1})\cdots S_{j,M}(-k_{j},k_M)
A^{(\ldots,-)},\no\\
A^{(\ldots,-)}&=&e^{2ik_jN}\bar{K}^-_j(k_j)\,A^{(\ldots,+)},\no\\
A^{(\ldots,+)}&=&S_{M,j}(k_M,k_j)\cdots S_{j+1,j}(k_{j+1},k_j)A^{(\ldots,+,\ldots)}.\no
\end{eqnarray}
Consequently, this gives rise to  the following eigenvalue problem:
\begin{eqnarray}
{\bar\tau}(k_j)A^{(\ldots,+,\ldots)}=e^{-2ik_jN}A^{(\ldots,+,\ldots)},\label{e}
\end{eqnarray}
with the resulting operators
\begin{eqnarray}
{\bar\tau}(k_j)&=&S_{j-1,j}(k_{j-1},k_j)\cdots S_{1,j}(k_{1},k_j){\bar K}_j^{+}(k_j)S_{j,1}(-k_j,k_1)\cdots
S_{j,j-1}(-k_j,k_{j-1})\no\\
&&\times S_{j,j+1}(-k_j,k_{j+1})\cdots S_{j,M}(-k_j,k_M) {\bar K}_j^-( k_j)\no\\
&&\times S_{M,j}(k_M,k_j)
\cdots S_{j+1,j}(k_{j+1},k_j).\label{H-operators}
\end{eqnarray}
Let ${\rm\bf V}$ denotes a two-dimensional linear space. Throughout the paper we adopt the standard
notations: for any matrix $A\in {\rm End}({\rm\bf V})$, $A_j$ is an
embedding operator in the tensor space ${\rm\bf V}\otimes
{\rm\bf V}\otimes\cdots$, which acts as $A$ on the $j$-th space and as
identity on the other factor spaces; $R_{ij}(u)$ is an embedding
operator of R-matrix in the tensor space, which acts as identity
on the factor spaces except for the $i$-th and $j$-th ones.

In the next section we shall show that ${\bar\tau}(k_j)$ is
proportional to the transfer matrix of the inhomogeneous XXX spin
chain with arbitrary boundary fields and thus the eigenvalue problem ({\ref e}) can be further
solved by the off-diagonal Bethe ansatz method.
\section{Off-diagonal Bethe ansatz}
\label{OBA} \setcounter{equation}{0}

Before going further, let us introduce the following R-matrix and K-matrices:
\begin{eqnarray}
R_{0,j}(u)&=&u+\eta\, P_{0,j},\\
K_0^-(u)&=&p+u\,{\vec h}_N\cdot{\vec \sigma}_0,\label{K-}\\
K_0^+(u)&=&q-(u+\eta)\,{\vec h}_1\cdot{\vec \sigma}_0,\label{K+}
\end{eqnarray}
where
\begin{eqnarray}
\eta=-i\frac U{2t},\quad p=i\frac{{\vec h}_N^2-t^2}{2t}, \quad
q=i\frac{t^2-{\vec h}_1^2}{2t}.\nonumber
\end{eqnarray}
The R-matrix possesses the following properties:
\begin{eqnarray}
&&\mbox{Initial condition}:\,R_{1,2}(0)= \eta P_{1,2},\label{Int-R}\\
&&\mbox{Unitarity relation}:\,R_{1,2}(u)R_{1,2}(-u)= -(u+\eta)(u-\eta)\,{\rm id},\label{Unitarity}\\
&&\mbox{Crossing relation}:\,R_{12}(u)=V_1R_{12}^{t_2}(-u-\eta)V_1,\quad V=-i\s^y.\label{crosing-unitarity}
\end{eqnarray}
The following Yang-Baxter equation, the reflection equation and its dual also hold:
\begin{eqnarray}
&&\hspace{-1.2truecm}R_{0,0'}(u-v)R_{0,1}(u)R_{0',1}(v)=R_{0',1}(v)R_{0,1}(u)R_{0,0'}(u-v),\label{YBeq}\\
&&\hspace{-1.2truecm}R_{0,0'}(u-v)K_0^-(u)R_{0,0'}(u+v)K_{0'}^-(v)=K_{0'}^-(v)R_{0,0'}(u+v)K_0^-(u)R_{0,0'}(u-v),\\
&&\hspace{-1.2truecm}R_{0,0'}(v\hspace{-0.1truecm}-\hspace{-0.1truecm}u){K_0^+}(u)
R_{0,0'}(\hspace{-0.08truecm}-\hspace{-0.08truecm}u\hspace{-0.08truecm}-\hspace{-0.08truecm}v-\hspace{-0.1truecm}2\eta){K_{0'}^+}(v)
\hspace{-0.12truecm}=\hspace{-0.12truecm}{K_{0'}^+}(v)
R_{0,0'}(\hspace{-0.08truecm}-\hspace{-0.08truecm}u\hspace{-0.08truecm}-\hspace{-0.08truecm}v\hspace{-0.08truecm}-\hspace{-0.08truecm}2\eta)
{K_0^+}(u)R_{0,0'}(v\hspace{-0.08truecm}-\hspace{-0.08truecm}u).
\end{eqnarray}
Now let us define the inhomogeneous double-row monodromy matrix \footnote{In order to compare with the operators (\ref{H-operators}), we choose
the inhomogeneous parameters $\theta_j=\sin k_j$.}  \cite{skl,Kor93},
\begin{eqnarray}
T_0(u)&=&R_{0,1}(u-\sin k_1)\cdots R_{0,M}(u-\sin
k_M)K_0^-(u)\nonumber\\
&&\times R_{M,0}(u+\sin k_M)\cdots
R_{1,0}(u+\sin k_1),
\end{eqnarray}
and the transfer matrix $\tau(u)$,
\begin{eqnarray}
\tau(u)=tr_0\lt\{{K}_0^+(u)T_0(u)\rt\}.\label{tau}
\end{eqnarray}
From the Yang-Baxter equation and the reflection equation and its dual one may derive \cite{skl}
\begin{eqnarray}
R_{0,0'}(u-v)T_0(u)R_{0,0'}(u+v)T_{0'}(v)=T_{0'}(v)R_{0,0'}(u+v)T_0(u)R_{0,0'}(u-v),
\end{eqnarray}
and the transfer matrices with different spectrum parameters commute with each other,
\begin{eqnarray}
[\tau(u), \tau(v)]=0.
\end{eqnarray}
Putting $u=-\sin k_j$, we readily have
\begin{eqnarray}
\tau(-\sin k_j)&=&R_{j-1,j}(-\sin k_j+\sin k_{j-1})\cdots R_{1,j}(-\sin k_j+\sin k_{1})\no\\
&&\times tr_0\lt\{ K_0^+(-\sin k_j)R_{0,j}(-2\sin k_j)R_{0,j}(0)\rt\}\nonumber\\
&&\times R_{j,1}(-\sin k_j-\sin k_1)\cdots R_{j,j-1}(-\sin k_j-\sin k_{j-1})\no\\
&&\times R_{j,j+1}(-\sin k_j-\sin k_{j+1})\cdots  R_{j,M}(-\sin k_j-\sin k_M)\nonumber\\
&&\times K_j^-(-\sin k_j)R_{M,j}(-\sin k_j+\sin k_M)\cdots R_{j+1,j}(-\sin k_j+\sin k_{j+1}).\no\\
\end{eqnarray}
Noticing that
\begin{eqnarray}
&&S_{j,l}(k_j,k_l)=\frac{R_{j,l}(\sin
k_j-\sin k_l)}{\sin k_j-\sin k_l+\eta},\\
&&S_{j,l}(-k_j,k_l)=\frac{R_{j,l}(-\sin
k_j-\sin k_l)}{-\sin k_j-\sin k_l+\eta},\\
&&{\bar K}_j^-(k_j)=\frac{2it\,K_j^-(-\sin
k_j)}{{\vec h}_N^2-t^2e^{-2ik_j}},\\
&&{\bar K}_j^+(k_j)=\frac{tr_0\lt\{it\,K_0^+(-\sin k_j)R_{0,j}(-2\sin
k_j)P_{0,j}\rt\}}{(\sin k_j-\eta)({\vec h}_1^2e^{2ik_j}-t^2)},
\end{eqnarray}
we have the following important identification between the operators
$\{{\bar \tau}(k_j)\}$ (\ref{H-operators}) appeared in the
eigenvalue problem of the open-boundary Hubbard model and the
transfer matrix of the open XXX spin chain with boundary fields:
\begin{eqnarray}
{\bar \tau}(k_j)&=&\prod_{l\neq j}^M(\sin k_j-\sin
k_l-\eta)^{-1}(\sin
k_j+\sin k_l-\eta)^{-1}\nonumber\\
&&\times \frac{-2t^2\,\tau(-\sin k_j)}{\eta(\sin
k_j-\eta)(t^2-{\vec h}_1^2e^{2ik_j})(t^2e^{-2ik_j}-{\vec h}_N^2)}.\label{tt}
\end{eqnarray}
The eigenvalue problem ({\ref e}) is thus equivalent to that of
diagonalizing the transfer matrix of the inhomogeneous open XXX chain model with boundary fields.
Here we naturally have the ``inhomogeneous" parameters
$\theta_j=\sin k_j$ and the crossing parameter $\eta=-i\frac{U}{2t}$. Thanks to the works \cite{cysw1,cysw2,cysw3},
the transfer matrix (\ref{tau}) of the open XXX chain with arbitrary boundary fields which is specified  by the K-matrices
$K^{\pm}(u)$ (\ref{K-}) and (\ref{K+}) can be exactly diagonalized by off-diagonal Bethe ansatz method. In the following, we
shall use the method in \cite{cysw3} to the eigenvalue problem (\ref{e}) of the Hubbard model with arbitrary boundary fields.

For this purpose, we introduce some functions at first:
\begin{eqnarray}
&&A(u)=\prod_{l=1}^M(u-\sin k_l+\eta)(u+\sin k_l+\eta),\label{A-function}\\
&&a(u)=\frac{2u+2\eta}{2u+\eta}(p+u\,{\rm sgn}(\vec h_1\cdot\vec h_N)|\vec h_N|)
(q-u\,|\vec h_1|) A(u),\label{a-function}\\
&&d(u)= a(-u-\eta),\label{d-function}\\
&& c=2({\rm sgn}(\vec h_1\cdot\vec h_N)|\vec h_1||\vec h_N|-\vec h_1\cdot \vec h_N)).\label{c-function}
\end{eqnarray}

\subsection{Even $M$ case}

Following \cite{cysw3}, we construct the following ansatz  of the eigenvalue of the transfer matrix $\tau(u)$ for an even $M$:
\begin{eqnarray}
\Lambda(u)= a(u)\frac{Q_1(u-\eta)}{Q_2(u)}+ d(u)\frac{Q_2(u+\eta)}{Q_1(u)}+ c\,u(u+\eta)\frac{ A(u) A(-u-\eta)}{Q_1(u)Q_2(u)},\label{T-Q-1}
\end{eqnarray}
in which the functions $Q_1(u)$ and $Q_2(u)$ are parameterized by $M$ different from each other parameters $\{\mu_j|j=1,\ldots,M\}$ for a generic
non-vanishing $c$ as follows:
\begin{eqnarray}
&&Q_1(u)=\prod_{j=1}^M(u-\mu_j),\\
&&Q_2(u)=\prod_{j=1}^M(u+\mu_j+\eta)=Q_1(-u-\eta).
\end{eqnarray} It has been shown \cite{cysw3} that $\Lambda(u)$ becomes the eigenvalue of the transfer matrix $\tau(u)$ given by (\ref{tau}) if the $M$ parameters
 $\{\mu_j|j=1,\ldots,M\}$ satisfies the following Bethe ansatz equations:
\begin{eqnarray}
&&\frac{c\,(\mu_j+\eta)(\mu_j+\frac{\eta}{2})}{(p-(\mu_j+\eta)\,{\rm sgn}(\vec h_1\cdot\vec h_N)|\vec h_N|)(q+(\mu_j+\eta)|\vec h_1|)}\no\\[6pt]
&&\quad=-\prod_{l=1}^M\frac{(\mu_j+\mu_l+\eta)(\mu_j+\mu_l+2\eta)}{(\mu_j-\sin k_l+\eta)(\mu_j+\sin k_l+\eta)},\quad j=1,\ldots,M,\label{BA-even-1}
\end{eqnarray} where the parameter $c$ is expressed in terms of the boundary fields (\ref{c-function}).  Numerical checks of the completeness of the above solutions for small size of $M$ (the results for the odd M see the next subsection) was given in \cite{Nep13-1,Jia13} (see also \cite{Nep13-2}).
A beautiful expression for the corresponding eigenvectors  was proposed recently in \cite{Bel13}.

Based on the expressions (\ref{T-Q-1}) of $\Lambda(u)$ for the eigenvalue of the transfer matrix (\ref{tau}) and
the relation (\ref{tt}) between  the operator $\bar\tau(k_j)$ and the transfer matrix at special point $\tau(-\sin k_j)$,
the eigenvalue problem (\ref{e}) gives rise to the following  constraints on  the quasi-momentum $\{k_j\}$:
\begin{eqnarray}
e^{-2ik_jN}&=&\prod_{l\neq j}^M(\sin k_j-\sin
k_l-\eta)^{-1}(\sin
k_j+\sin k_l-\eta)^{-1}\nonumber\\
&&\times\frac{-2t^2\Lambda(-\sin k_j)}{\eta(\sin
k_j-\eta)(t^2-{\vec h}_1^2e^{2ik_j})(t^2e^{-2ik_j}-{\vec h}_N^2)}.
\end{eqnarray}
Noticing that $d(-\sin k_j)=A(\sin k_j-\eta)=0$, the above Bethe ansatz equations  become
\begin{eqnarray}
&&\frac{4t^2(p-\sin k_j\,{\rm sgn}(\vec h_1\cdot\vec h_N)|\vec h_N|)(q+\sin k_j\,|\vec h_1|)}{(t^2-\vec h_1^2e^{2ik_j})(t^2e^{-2ik_j}-\vec h_N^2)}=e^{-2ik_jN}\,\prod_{l=1}^M\frac{(\sin k_j-\mu_l-\eta)}{(\sin k_j+\mu_l+\eta)},\no\\
&&\quad j=1,\ldots,M.\label{BA-Even-2}
\end{eqnarray}
Then from the solutions of the Bethe ansatz equations (\ref{BA-even-1}) and (\ref{BA-Even-2}), one can reconstruct the exact wave
functions (\ref{Wave-function}) with even number of electrons for the Hubbard model with boundary fields, the corresponding eigenvalues
are given by (\ref{Eigenvalues}).

\subsection{Odd $M$ case}

Following \cite{cysw3}, we construct the following ansatz  of the eigenvalue of the transfer matrix $\tau(u)$ for an odd $M$:
\begin{eqnarray}
\Lambda(u)= a(u)\frac{Q_1(u-\eta)}{Q_2(u)}+ d(u)\frac{Q_2(u+\eta)}{Q_1(u)}+ c\,u^2(u+\eta)^2\frac{A(u) A(-u-\eta)}{Q_1(u)Q_2(u)},\label{T-Q}
\end{eqnarray}
where the functions $a(u)$, $d(u)$ and $A(u)$ and the parameter $c$ are given by (\ref{A-function})-(\ref{c-function}) respectively.
The functions $Q_1(u)$ and $Q_2(u)$ are some functions parameterized by $M+1$ different from each other parameters
$\{\mu_j|j=1,\ldots,M+1\}$ for a generic non-vanishing $c$ as follows:
\begin{eqnarray}
&&Q_1(u)=\prod_{j=1}^{M+1}(u-\mu_j),\\
&&Q_2(u)=\prod_{j=1}^{M+1}(u+\mu_j+\eta)=Q_1(-u-\eta).
\end{eqnarray}
Keeping the expression (\ref{T-Q}) of the function $\Lambda(u)$ in mind, we find that the $M$ quasi-momentum $\{k_j\}$ and the $M+1$ parameters
$\{\mu_j|j=1,\ldots,M+1\}$ need to satisfy the following Bethe ansatz equations:
\begin{eqnarray}
&&\frac{4t^2(p-\sin k_j\,{\rm sgn}(\vec h_1\cdot\vec h_N)|\vec h_N|)(q+\sin k_j\,|\vec h_1|)}{(t^2-\vec h_1^2e^{2ik_j})(t^2e^{-2ik_j}-\vec h_N^2)}=e^{-2ik_jN}\,\prod_{l=1}^{M+1}\frac{(\sin k_j-\mu_l-\eta)}{(\sin k_j+\mu_l+\eta)},\no\\
&&\quad j=1,\ldots,M,\label{BA-odd-1}\\[6pt]
&&\frac{ -c\,\mu_j(\mu_j+\frac{\eta}{2})(\mu_j+\eta)^2}{(p-(\mu_j+\eta)\,{\rm sgn}(\vec h_1\cdot\vec h_N)|\vec h_N|)(q+(\mu_j+\eta)|\vec h_1|)}\prod_{l=1}^{M}(\mu_j-\sin k_l+\eta)(\mu_j+\sin k_l+\eta)\no\\[6pt]
&&\quad=\prod_{l=1}^{M+1}(\mu_j+\mu_l+\eta)(\mu_j+\mu_l+2\eta),\quad j=1,\ldots,M+1.\label{BA-odd-2}
\end{eqnarray}
From the solutions of the Bethe ansatz equations (\ref{BA-odd-1}) and (\ref{BA-odd-2}), one can reconstruct the exact wave
functions (\ref{Wave-function}) with odd number of electrons for the Hubbard model with boundary fields, the corresponding eigenvalues
are given by (\ref{Eigenvalues}).

\section{Reduction to the parallel boundary case}
\label{RED} \setcounter{equation}{0}

When the two boundary fields $\vec h_1$ and $\vec h_N$ are parallel or anti-parallel,
the $U(1)$ symmetry in the spin sector is recovered, and the associated open XXX spin chain is specified by two diagonal K-matrices. In our method the corresponding parameter $c$ given by (\ref{c-function}) is vanishing. The resulting $T-Q$ ansatz of the eigenvalue of the transfer matrix of the associated spin chain  reduces to the usual form no matter $M$ is even or odd \cite{cysw3}:
\begin{eqnarray}
\Lambda(u)= a(u)\frac{Q(u-\eta)}{Q(u)}+ d(u)\frac{Q(u+\eta)}{Q(u)},
\end{eqnarray}
where the functions $Q(u)$ are parameterized by $m$ different from each other Bethe roots $\{\l_j|j=1,\ldots,m\}$ with discrete $m=0,\ldots,M$ as follows:
\begin{eqnarray}
Q(u)=\prod_{l=1}^m(u-\l_l)(u+\l_l+\eta)=Q(-u-\eta).
\end{eqnarray} Here the discrete number $m$ is the consequence of the $U(1)$ symmetry reservation in the case that
the two boundary fields $\vec h_1$ and $\vec h_N$ are parallel or anti-parallel.
These $m$ parameters $\{\l_j\}$ and $M$ quasi-momentum $\{k_j\}$ satisfy the following Bethe ansatz equations:
\begin{eqnarray}
&&\frac{4t^2(p-\sin k_j|\vec h_N|)(q+\sin k_j|\vec h_1|)}{(t^2-\vec h_1^2e^{2ik_j})(t^2e^{-2ik_j}-\vec h_N^2)}=e^{-2ik_jN}\prod_{l=1}^m\frac{(\sin k_j+\l_l)(\sin k_j-\l_l-\eta)}{(\sin k_j-\l_l)(\sin k_j+\l_l+\eta)},\no\\
&&\quad j=1,\ldots,M,\label{BA-Para-1}\\[6pt]
&&\frac{\l_j(p-(\l_j+\eta)|\vec h_N|)(q+(\l_j+\eta)|\vec h_1|)}
{(\l_j+\eta)(p+\l_j|\vec h_N|)(q-\l_j|\vec h_1|)}\prod_{l=1}^M\frac{(\l_j+\sin k_l)(\l_j-\sin k_l)}{(\l_j-\sin k_l+\eta)(\l_j+\sin k_l+\eta)}\no\\[6pt]
&&\quad=-\prod_{l=1}^m\frac{(\l_j-\l_l-\eta)(\l_j+\l_l)}{(\l_j-\l_l+\eta)(\l_j+\l_l+2\eta)},\quad j=1,\ldots,m.\label{BA-Para-2}
\end{eqnarray}
From the solutions of the Bethe ansatz equations (\ref{BA-Para-1}) and (\ref{BA-Para-2}), one can reconstruct the exact wave
functions (\ref{Wave-function}) for the Hubbard model with parallel or anti-parallel boundary fields, the corresponding eigenvalues
are given by (\ref{Eigenvalues}).

\section{Conclusion}

The one-dimensional Hubbard model with arbitrary boundary magnetic fields described by the Hamiltonian
(\ref{Hamiltonian}) is studied by combining the coordinate Bethe ansatz and off-diagonal Bethe ansatz methods.
With the coordinate Bethe ansatz, eigen-functions of the Hamiltonian of the model are
given in terms of some quasi-momentum $\{k_j\}$ as (\ref{Wave-function}). The constraints (\ref{e}) on these quasi-momentum
is transformed into the eigenvalues problem of the resulting transfer matrix of the associated open XXX spin chain with arbitrary boundary
fields. The second eigenvalue problem is then solved via the off-diagonal Bethe ansatz method. The corresponding Bethe ansatz equations
(\ref{BA-even-1}) and (\ref{BA-Even-2}) for the even number of electrons case, (\ref{BA-odd-1}) and (\ref{BA-odd-2}) for the odd number of
electrons are constructed respectively  when two boundary fields are unparallel, which corresponds to the case of the $U(1)$ symmetry
in the spin sector being broken. When the two boundary fields $\vec h_1$ and $\vec h_N$ are parallel or anti-parallel, the $U(1)$ symmetry
in spin sector is recovered, the resulting Bethe ansatz equations become   (\ref{BA-Para-1}) and (\ref{BA-Para-2}) which are labeled by a
discrete number $m=0,\ldots,M$.

\section*{Acknowledgments}

Y. Wang would like to thank X.W. Guan for fruitful discussions. The
financial support from the National Natural Science Foundation of
China (Grant Nos. 11174335, 11075126, 11031005, 11375141, 11374334),
the National Program for Basic Research of MOST (973 project under
grant No.2011CB921700), the State Education Ministry of China
(Grant No. 20116101110017) and BCMIIS are gratefully
acknowledged.

\end{document}